\documentclass[12pt]{article}
\usepackage{epsfig}

\newcommand\pubnumber{MSUHEP-10615}
\newcommand\pubdate{\today}
\newcommand\hepnumber{hep-ph/0106181}
\def\lsim{\mathrel{\raise.3ex\hbox{$<$\kern-.75em\lower1ex\hbox{$\sim$}}}}
\def\gsim{\mathrel{\raise.3ex\hbox{$>$\kern-.75em\lower1ex\hbox{$\sim$}}}}

\def\csumb{Department of Physics and Astronomy\\
Michigan State University, East Lansing, MI 48824 USA}
\def\support{\footnote{Work supported by the
US National Science Foundation under grants PHY-9722144
and PHY-0070443.}} 

\def\Title#1{\begin{center} {\Large\bf #1 } \end{center}}
\def\Author#1{\begin{center}{ \sc #1} \end{center}}
\def\Address#1{\begin{center}{ \it #1} \end{center}}

\newcommand\pubblock{\rightline{\begin{tabular}{l} \pubnumber\\
         \pubdate\\ \hepnumber \end{tabular}}}
\newenvironment{Abstract}{\begin{quotation}  }{\end{quotation}}
\newenvironment{Presented}{\begin{quotation} \begin{center} 
             Presented at the\end{center}
      \begin{center}\begin{large}}{\end{large}\end{center} \end{quotation}}

\makeatletter
\def\section{\@startsection{section}{0}{\z@}{5.5ex plus .5ex minus
 1.5ex}{2.3ex plus .2ex}{\large\bf}}
\def\subsection{\@startsection{subsection}{1}{\z@}{3.5ex plus .5ex minus
 1.5ex}{1.3ex plus .2ex}{\normalsize\bf}}
\def\subsubsection{\@startsection{subsubsection}{2}{\z@}{-3.5ex plus
-1ex minus  -.2ex}{2.3ex plus .2ex}{\normalsize\sl}}

\renewcommand{\@makecaption}[2]{%
   \vskip 10pt
   \setbox\@tempboxa\hbox{\small #1: #2}
   \ifdim \wd\@tempboxa >\hsize     
       \small #1: #2\par          
     \else                        
       \hbox to\hsize{\hfil\box\@tempboxa\hfil}
   \fi}

 \def\citenum#1{{\def\@cite##1##2{##1}\cite{#1}}}
 
\newcount\@tempcntc
\def\@citex[#1]#2{\if@filesw\immediate\write\@auxout{\string\citation{#2}}\fi
  \@tempcnta\z@\@tempcntb\m@ne\def\@citea{}\@cite{\@for\@citeb:=#2\do
    {\@ifundefined
       {b@\@citeb}{\@citeo\@tempcntb\m@ne\@citea\def\@citea{,}{\bf ?}\@warning
       {Citation `\@citeb' on page \thepage \space undefined}}%
    {\setbox\z@\hbox{\global\@tempcntc0\csname b@\@citeb\endcsname\relax}%
     \ifnum\@tempcntc=\z@ \@citeo\@tempcntb\m@ne
       \@citea\def\@citea{,}\hbox{\csname b@\@citeb\endcsname}%
     \else
      \advance\@tempcntb\@ne
      \ifnum\@tempcntb=\@tempcntc
      \else\advance\@tempcntb\m@ne\@citeo
      \@tempcnta\@tempcntc\@tempcntb\@tempcntc\fi\fi}}\@citeo}{#1}}
\def\@citeo{\ifnum\@tempcnta>\@tempcntb\else\@citea\def\@citea{,}%
  \ifnum\@tempcnta=\@tempcntb\the\@tempcnta\else
  {\advance\@tempcnta\@ne\ifnum\@tempcnta=\@tempcntb \else\def\@citea{--}\fi
    \advance\@tempcnta\m@ne\the\@tempcnta\@citea\the\@tempcntb}\fi\fi}
\makeatother

\begin{document}
\begin{titlepage}
\pubblock

\vfill
\def\thefootnote{\fnsymbol{footnote}}
\Title{Review of BFKL}
\vfill
\Author{Carl R. Schmidt\support}
\Address{\csumb}
\vfill
\begin{Abstract}
I describe the underlying physics behind the BFKL resummation
and discuss some of the recent ideas and results in this field.
On the theoretical side I consider the formalism in the next-to-leading
logarithmic (NLL) approximation and the interpretation of the
large corrections.  On the phenomenological side I discuss several
experiments that attempt to observe the BFKL effects in the high energy
limit.
\end{Abstract}
\vfill
\begin{Presented}
5th International Symposium on Radiative Corrections \\ 
(RADCOR--2000) \\[4pt]
Carmel CA, USA, 11--15 September, 2000
\end{Presented}
\vfill
\end{titlepage}
\def\thefootnote{\arabic{footnote}}
\setcounter{footnote}{0}

\section{Introduction to BFKL}      

Hadronic processes at high-energy colliders often involve more
than one energy scale.  As a consequence, calculations in 
perturbative Quantum Chromodynamics (QCD) can involve large 
logarithms of the ratio of these scales,
which must be resummed to obtain a reliable prediction.
Two processes where this might be necessary are
Deep Inelastic Scattering (DIS) at small $x$ and hadronic
dijet production at large rapidity intervals $\Delta y$.  
In DIS the logarithm that 
appears is $\ln(1/x)$, with $x \simeq Q^2/s$ 
the squared ratio of the momentum transfer to the photon-hadron 
center-of-mass energy.  In large-rapidity dijet production the large 
logarithm is $\Delta y\simeq\ln(\hat s/|\hat t|)$, with
$\hat s$ the squared parton center-of-mass energy and $|\hat t|$ of the
order of the squared jet transverse energy.  In both of these cases
the large logarithms, which arise at each order in the coupling 
constant $\alpha_{s}$, can be resummed  by means of the
Balitsky-Fadin-Kuraev-Lipatov (BFKL) equation \cite{FKL}.

The most familiar prediction of the BFKL resummation in the leading
logarithmic (LL) approximation is the power-law rise in the partonic
cross section as a function of the energy:
\begin{equation}
	\hat\sigma\ \approx e^{A\Delta y} \approx\ 
	{\hat s}^{A}\ .
	\label{saddle}
\end{equation}
The quantity $(1+A)$ is often referred to as the BFKL Pomeron 
intercept, where at LL
\begin{equation}
       A\ =\ {\bar\alpha_s4\ln2}
\label{Avalue}
\end{equation}
with $\bar\alpha_s=N_c\alpha_s/\pi$ and $N_c=3$ the number of colors.  
Similarly, in DIS the structure 
functions are predicted to rise as $x^{-A}$ at small $x$.
One of the goals in BFKL physics has been to observe this power-law 
rise as a direct indication of the importance of the resummation.
In this talk I will review the status of BFKL physics, both in its
theoretical development at next-to-leading logarithm (NLL) and in its
phenomenological application to experiment.  

\section{BFKL at LL}
\label{bfklatll}

I begin by giving a simple physical picture of the BFKL
resummation at leading logarithm (LL), 
where the large logarithm is taken to be
the rapidity interval between two widely-separated partonic jets.
Although this in no way can be considered a derivation of the
BFKL equation, it is useful to show what assumptions go into the
resummation and to show how the factors of $\alpha_s\Delta y$
arise at each order and exponentiate.

\begin{figure}[t]	
\centerline{\epsfxsize 4.0 truein \epsfbox{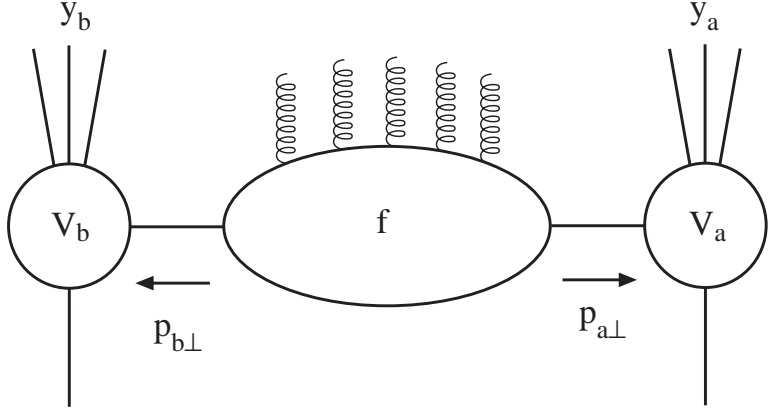}}
\vskip 0.0 cm
\caption[]{
\label{factorize}
\small 
 A schematic picture of the cross section for producing particles
at large rapidity separation.}
\end{figure}

The starting point is the factorization of the
partonic cross section at large rapidity separation:
\begin{equation}
{d\hat\sigma\over d^2p_{a\perp}
d^2p_{b\perp}}\ =\ V_a(p_{a\perp}^2)\,
f(\Delta y,p_{a\perp},p_{b\perp})\, V_b(p_{b\perp}^2) \label{semihard}\ .
\end{equation}
A physical interpretation of this factorization is represented
in Fig.~\ref{factorize}.   The process consists of two distinct scatterings,
 which occur at widely-separated rapidities, $y_a$ and $y_b$,
and small transverse momenta $|p_{a\perp}|\sim |p_{b\perp}|$.
The impact factors $V(p_\perp^2)$ depend only on the transverse momentum
exchanged and the specific partons involved in each scattering, but not
on anything else that happens in the event.  The function $f(\Delta y,
p_{a\perp},p_{b\perp})$ connects the two scatterings by accounting
for the emission of gluons in the rapidity interval.  This function
is universal and provides the exponentiation of the logarithms.

\begin{figure}[t]	
\centerline{\epsfysize 1.3 truein \epsfbox{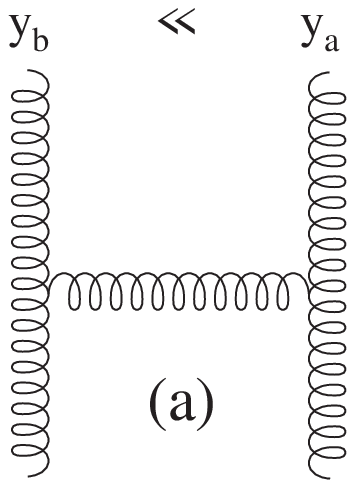}\hskip0.7in
\epsfysize 1.3 truein \epsfbox{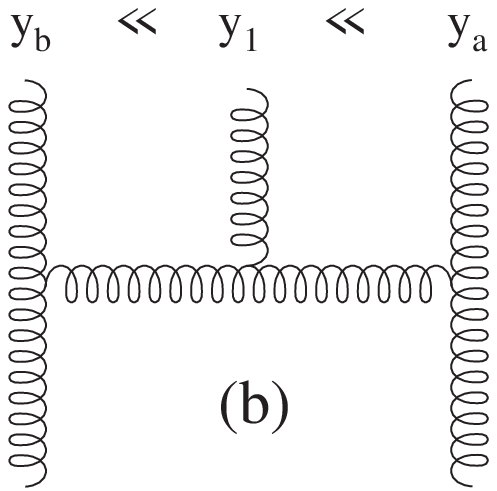}\hskip0.7in
\epsfysize 1.3 truein \epsfbox{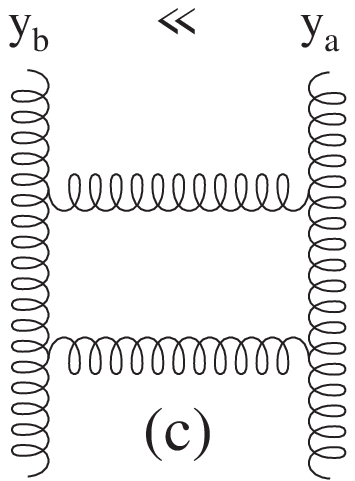}}   
\vskip -.0 cm
\caption[]{
\label{llladder}
\small Contributions to LL BFKL ladder obtained from $gg$ scattering:\\ 
(a) $(\alpha_s\Delta y)^0$ real.
(b) $(\alpha_s\Delta y)^1$ real.
(c) $(\alpha_s\Delta y)^1$ virtual.}
\end{figure}

We can see this factorization directly in the Born cross section for
gluon-gluon scattering at $y_a \gg y_b$, shown diagrammatically 
in Fig.~\ref{llladder}(a):
\begin{equation}
{d{\hat\sigma}_{gg}^{(0)}\over d^2p_{a\perp}d^2p_{b\perp}}\ =\ 
\left[{N_c\alpha_s\over p_{a\perp}^2}\right]
\left[{1\over2}\delta^{(2)}(p_{a\perp}+p_{b\perp})\right]
\left[{N_c\alpha_s\over p_{b\perp}^2}\right]\ ,
\label{born}
\end{equation}
where the central factor is the ${\cal O}(\alpha_s^0)$ contribution to the
function $f$, and we see that the leading-order gluon impact factor is
\begin{equation}
V_g(p_\perp^2)\ =\ \frac{N_c\alpha_s}{p_\perp^2}\ .
\end{equation}
The real ${\cal O}(\alpha_s^1)$ correction to $f$ can
be obtained by considering the emission of three gluons, strongly ordered
in rapidity $y_a\gg y_1\gg y_b$, shown diagrammatically 
in Fig.~\ref{llladder}(b):
\begin{equation}
{d{\hat\sigma}_{gg}^{(1r)}\over d^2p_{a\perp}d^2p_{b\perp}}\ =\ 
\left[{N_c\alpha_s\over p_{a\perp}^2}\right]
\left[{{\bar\alpha_s}\over\pi}\int{d^2k_{1\perp}dy_1\over k_{1\perp}^2}
\,{1\over2}\delta^{(2)}(p_{a\perp}+k_{1\perp}+p_{b\perp})\right]
\left[{N_c\alpha_s\over p_{b\perp}^2}\right]\ .
\label{realone}
\end{equation}
{}From this formula, we easily see where the large logarithm comes from.  
It arises from the integral over the rapidity $y_1$ of the intermediate
gluon, resulting in a factor of $\alpha_s\Delta y$.

We can now generalize the form of these real corrections to an arbitrary
number of emitted gluons:
\begin{itemize}
\item At each order, another gluon is inserted in the ladder
with a weight given by
\begin{eqnarray}
&&\frac{\bar\alpha_s}{\pi}\frac{d^2k_{i\perp}}{k_{i\perp}^2}\ .\nonumber
\end{eqnarray}
\item The emitted gluons conserve transverse momentum, as enforced by a
delta-function factor
\begin{eqnarray}
&&\frac{1}{2}\delta^{(2)}(p_{a\perp}+\sum_i k_{i\perp}+p_{b\perp})\ .\nonumber
\end{eqnarray}
\item The intermediate gluons are integrated over the ordered rapidity-intervals,
\begin{eqnarray}
&&y_b<y_1<y_2<\cdots<y_n<y_a\ ,\nonumber
\end{eqnarray}
producing an overall factor of
\begin{eqnarray}
&&\frac{(\Delta y)^n}{n!}\ .\nonumber
\end{eqnarray}
\end{itemize}
The real $n$-gluon contribution can be obtained directly from
the tree-level $(n+2)$-gluon cross-section by assuming that all
transverse momenta are comparable in size,
while terms suppressed by ${\cal O}(e^{-|y_i-y_j|})$
are neglected.

Of course, the real contributions by themselves are not infrared-safe,
because they diverge when the $k_\perp$ of any of the intermediate gluons 
vanishes.  The cure for this is the inclusion of virtual corrections
in the large $\Delta y$ limit.  The ${\cal O}(\alpha_s^1)$ virtual
correction in $gg$ scattering 
comes from the diagram in Fig.~\ref{llladder}(c).  
Similar contributions
must come in at each order to regularize the $1/k_\perp^{2}$ 
infrared singularities.
Including these corrections, we can understand the following form of
the BFKL equation
\begin{eqnarray}
f(\Delta y,p_{a\perp},p_{b\perp})
&=&\frac{1}{2}\delta^{(2)}(p_{a\perp}+p_{b\perp})\nonumber\\
&&+(\Delta y)
K\biggl[\frac{1}{2}\delta^{(2)}(p_{a\perp}+p_{b\perp})\biggr]\nonumber\\
&&+\frac{1}{2!}(\Delta y)^2
K\biggl[K\biggl[\frac{1}{2}\delta^{(2)}(p_{a\perp}+p_{b\perp})\biggr]\biggr]\nonumber\\
&&+\frac{1}{3!}(\Delta y)^3K\biggl[
K\biggl[K\biggl[\frac{1}{2}\delta^{(2)}(p_{a\perp}+p_{b\perp})\biggr]\biggr]\biggr]\nonumber\\
&&+\cdots\ ,
\label{llbfkl}
\end{eqnarray}
where the kernel $K$ is an 
integral operator acting on a function $\phi(p_{a\perp})$ by 
\begin{equation}
K\biggl[\phi(p_{a\perp})\biggr]\ =\ \frac{\bar\alpha_s}{\pi}
\int \frac{d^2k_\perp}{k_\perp^2}\left[
\phi(k_\perp+p_{a\perp})-\frac{p_{a\perp}^2}{k_\perp^2+(k_\perp
+p_{a\perp})^2}\phi(p_{a\perp})\right]\ .
\label{kernel}
\end{equation}
At LL, each operation of the kernel inserts one more real or virtual gluon 
in the ladder.  The first term in (\ref{kernel}) corresponds to a real gluon, 
while the second term
is the virtual correction needed to regularize the soft singularity.
The equation (\ref{llbfkl}) is just the solution of the BFKL
equation obtained by iteration.  

A more compact form of the BFKL solution at LL is obtained by finding
the eigenfunctions and eigenvalues of the kernel (\ref{kernel}).
This yields
\begin{equation}
f(\Delta y,p_{a\perp},p_{b\perp})\ =\ {1\over i\,(2\pi)^2 }
\sum_{n=-\infty}^{\infty}e^{in\tilde\phi}
\int_{1/2-i\infty}^{1/2+i\infty} d\gamma\, 
(p_{a\perp}^2)^{\gamma-1}\,(p_{b\perp}^{2})^{-\gamma}
e^{\bar\alpha_s\chi(n,\gamma)\Delta y}\ ,\label{sold}
\end{equation}
where $\tilde\phi=\phi_{a}-\phi_b-\pi$, the function
\begin{equation}
	\bar\alpha_s\chi(n,\gamma)\ =\ \bar\alpha_s\left[2\psi(1)
-\psi({\textstyle\frac{n}{2}}+\gamma)
-\psi({\textstyle\frac{n}{2}}+1-\gamma)\right]
	\label{eig}
\end{equation}
gives the eigenvalues of the LL BFKL kernel (\ref{kernel}), and $\psi$ is the  
logarithmic derivative of the gamma function.  
For very large $\Delta y$, 
the integral over $\gamma$ in (\ref{sold}) can be performed in
the saddle-point approximation.  The $n=0$ term dominates,
and one obtains the exponential 
rise in the cross section (\ref{saddle}), displayed in the introduction 
with $A=\bar\alpha_s\chi(0,\frac{1}{2})=4\bar\alpha_{s}\ln2$. 

\section{BFKL at NLL}

\begin{figure}[t]	
\centerline{\epsfysize 1.3 truein \epsfbox{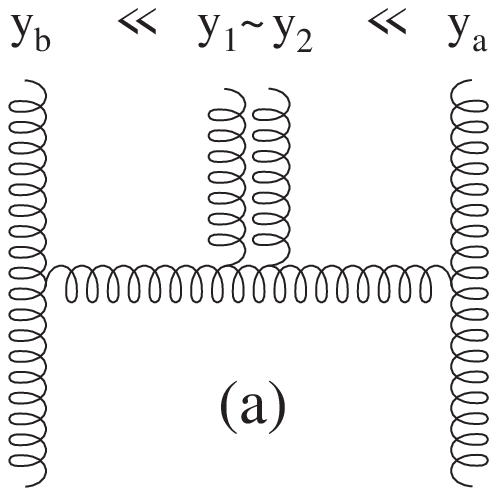}\hskip0.7in
\epsfysize 1.3 truein \epsfbox{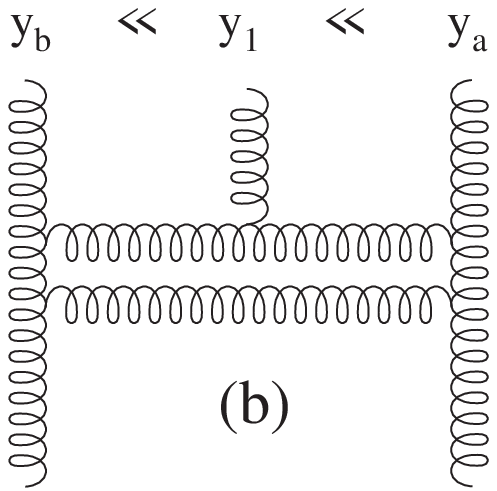}\hskip0.7in
\epsfysize 1.3 truein \epsfbox{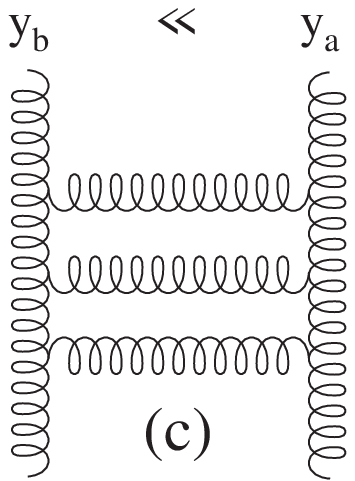}}   
\vskip -.0 cm
\caption[]{
\label{nllladder}
\small Contributions to NLL BFKL ladder at ${\cal O}(\alpha_s^2\Delta y)$
obtained from $gg$ scattering.}
\end{figure}

At LL each application of the kernel gives a contribution
of ${\cal O}(\alpha_s\Delta y)$.  At NLL one also includes terms of 
${\cal O}(\alpha_{s}^{2}\Delta y)$.
That is, we can reinterpret the kernel in (\ref{llbfkl}) as a
power series $K\equiv \bar\alpha_s K^{(1)}+\bar\alpha_s^2 K^{(2)}+\dots$, 
where $\bar\alpha_s K^{(1)}$
is the LL kernel, given in (\ref{kernel}), and $\bar\alpha_s^2K^{(2)}$ includes
the NLL corrections.
There are three types of contributions at NLL, which
are shown schematically, in the context of $gg$ scattering, in 
Fig.~\ref{nllladder}.  They consist of:
(a) the emission of 
two gluons nearby in rapidity, 
(b) the virtual correction to the emission
of one gluon, widely separated in rapidity, (c) the subleading 
purely-virtual corrections.  
These three types of contributions took many 
years and many papers to sort out the technical details\footnote{A 
list of references can be found in ref.~\cite{schmidt}, but with no 
guarantee of completeness.}, with the final NLL kernel obtained
in 1998 by Fadin and Lipatov \cite{nll}.  
  Although the full kernel has not been 
checked in a completely independent manner, many of the pieces of the 
calculation have received independent confirmation.  Two particularly 
significant checks are the 
calculation of the virtual correction to the gluon emission at large
rapidity separation \cite{ddsii}, and the compilation of the three 
NLL terms into a single kernel with the cancellation of all collinear and 
soft singularities \cite{cc}.

The final result of this calculation is usually presented by applying 
the NLL kernel to the LL eigenfunctions, with azimuthal averaging,  
yielding
\begin{eqnarray}
	K_{\rm NLL}\left[
	(p_{a\perp}^{2})^{\gamma-1}\right]\ &=& \ \
	\biggl\{\bar\alpha_s(\mu)\chi(\gamma)\biggr\}\,
        (p_{a\perp}^{2})^{\gamma-1}\nonumber\\
	&=&\ 
	\biggl\{\bar\alpha_s(\mu)
	\chi^{(1)}(\gamma)\left[
	1-\bar\alpha_{s}(\mu)b_{0}
	\ln(p_{a\perp}^{2}/\mu^{2})\right]\nonumber\\
        &&
	+\ \bar\alpha_{s}(\mu)^2\chi^{(2)}(\gamma)\biggr\}\,
        (p_{a\perp}^{2})^{\gamma-1}
	\label{nll}
\end{eqnarray}
where $\bar\alpha_s\chi^{(1)}(\gamma)$ is the LL eigenvalue for $n=0$ 
given in (\ref{eig}),
$b_{0}= {11/12}-{n_{f}/(6N_{c})}$, and we have explicitly included
the dependence on the $\overline{\rm MS}$ renormalization scale $\mu$.  
The NLL correction has been separated into two terms.  The first term 
depends on the scale $p_{a\perp}$ and is associated with the running 
of the coupling in the LL kernel:  
$\alpha_{s}(\mu)\rightarrow \alpha_{s}(p_{a\perp})$.  
The second term, $\bar\alpha_{s}^{2}\chi^{(2)}(\gamma)$, is independent 
of scale and contains the remainder of the NLL corrections \cite{nll}.


After completion of the NLL corrections to the BFKL kernel, several 
issues quickly became apparent.  
Roughly speaking, they can be separated into 
issues associated with the running coupling term and issues associated 
with the scale-invariant term.  In this talk I will concentrate on the 
scale-invariant term.  For analyses of the running coupling issues, 
see Refs.~\cite{Kovchegov,Armesto,levin,CCS}. 

\begin{figure}[t]	
\centerline{\epsfxsize 3.5 truein \epsfbox{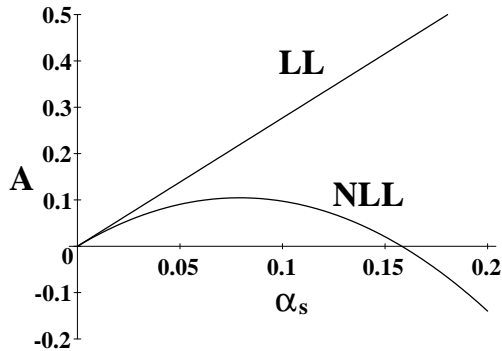}}
\vskip -1.0 cm
\caption[]{
\label{nlleig}
\small 
Leading BFKL eigenvalue $A=\bar\alpha_s\chi(\frac{1}{2})$ at LL and NLL.}
\end{figure}

The first indication of problems with BFKL at NLL was 
seen immediately by Fadin and Lipatov.  The corrections to the 
leading eigenvalue  are large and negative!  If we ignore the 
effects of running coupling, we obtain 
\begin{equation}
	\bar\alpha_{s}\chi(\mbox{$1\over2$})\ =\ 
	2.77\bar\alpha_{s}-18.34\bar\alpha_{s}^{2}\ ,
	\label{nllchi}
\end{equation}
for three active flavors.  
This function is plotted in Fig.~\ref{nlleig}.
At the not-unreasonable value 
of $\alpha_{s}=0.16$ the NLL corrections exactly cancel the LL term, 
while for larger values of $\alpha_{s}$ the eigenvalue becomes negative.
Naively, this would indicate that the BFKL Pomeron intercept also 
becomes negative, leading to a cross section that decreases, rather 
than increases, as a power of the energy.

Unfortunately, things get even worse.  The standard BFKL
power-law scaling of the partonic cross section (\ref{saddle})
relies on the saddle-point evaluation 
of the NLL generalization of the BFKL solution (\ref{sold}).  Upon 
closer analysis Ross \cite{ross} showed that the NLL eigenvalue 
function $\chi(\gamma)$ no longer has a maximum at $\gamma={1\over2}$, but 
has a minimum with two maxima occuring symmetrically on either side 
of this point\footnote{The standard procedure in these analyses is to 
modify the LL eigenfunctions used in eq.~(\ref{nll}) in order to make the
eigenvalues manifestly symmetric under $\gamma\rightarrow1-\gamma$, 
following
ref.~\cite{nll}.}.  Performing a higher-order expansion of $\chi(\gamma)$, 
Ross found a smaller correction to the BFKL Pomeron intercept.
However, the cross section he obtained was not positive definite.  It 
contained oscillations as one varied $p_{a\perp}$ and $p_{b\perp}$.
This led Levin \cite{levin} to declare that NLL BFKL has a serious 
pathology.

One might wonder whether the approximate evaluation of the integral 
performed by Ross is adequate at this stage.  Perhaps an exact evaluation 
is necessary.  However, negative cross sections have also arisen when 
the resummed small-$x$ anomalous dimensions, obtained from the NLL BFKL 
solution, were used to study DIS scattering at 
small-$x$ \cite{ball,blumlein}.  In any event the NLL corrections to the 
BFKL solution are large, leading one to question the stability and 
applicability of the BFKL resummation procedure in general.

\section{Understanding the large NLL Corrections.}
\label{sec:nllbig}

When any perturbation expansion has large corrections at higher
orders, the natural thing to do is to try to reorganize the series
so that it converges more rapidly.  In this talk I will briefly
discuss three different approaches to this reorganization.

The first proposal by Brodsky {\it et al.} \cite{blm}
uses the freedom to choose the renormalization scheme.
The NLL eigenvalue equation (\ref{nllchi}) is written in the
$\overline{\rm MS}$ scheme.  Brodsky {\it et al.} argued
that a non-abelian physical scheme should be more natural
for the BFKL resummation.  Then they used the BLM procedure 
\cite{blmone} to find the optimal scale for the QCD coupling.
In this case the BLM procedure dictates a large scale, thereby
reducing the effective $\alpha_s$ (and the LL prediction) and also 
reducing the coefficient of the NLL $\alpha_s^2$ term.
This approach predicts a BFKL Pomeron intercept of $A\sim0.17$
for $\alpha_s=0.2$.  In addition it yields a very weak 
dependence on the gluon virtuality $p_{a\perp}^{2}$ and leads to an 
approximate conformal invariance.

The motivation for the second proposal \cite{schmidt} (first suggested in 
\cite{bo} and \cite{liptalk}) can be seen from the discussion of the 
physics of BFKL at
LL in section \ref{bfklatll}.  The approximation used in deriving
the LL contribution at each order in the BFKL ladder was to neglect
terms of ${\cal O}(e^{-|y_i-y_{i+1}|})$ in the QCD matrix elements, which is 
valid when the emitted gluons are all widely separated in rapidity.
However, the gluon rapidity $y_{i}$ is then
integrated all the way up to $y_{i+1}$. 
Thus, the errors in the matrix elements 
are largest when $y_{i}\sim y_{i+1}$.   This 
suggests that one enforce a condition
$y_{i+1}-y_{i}>\Delta$, so that the gluons are required to be widely 
separated, and the kinematic approximations are good.  The arbitrary 
parameter $\Delta$ is assumed to be much smaller than the total rapidity 
interval.   
The excluded region is re-introduced at NLL, such that the change in the
cross section due to shifting $\Delta$ is always
next-to-next-to-leading logarithm (NNLL).  
In this way, the dependence on $\Delta$ can be regarded as an estimate 
of the uncertainty due to NNLL corrections (similar to the role of the
renormalization scale $\mu$ in the $\overline{\rm MS}$ scheme).

\begin{figure}[t]	
\centerline{\epsfxsize 4.0 truein \epsfbox{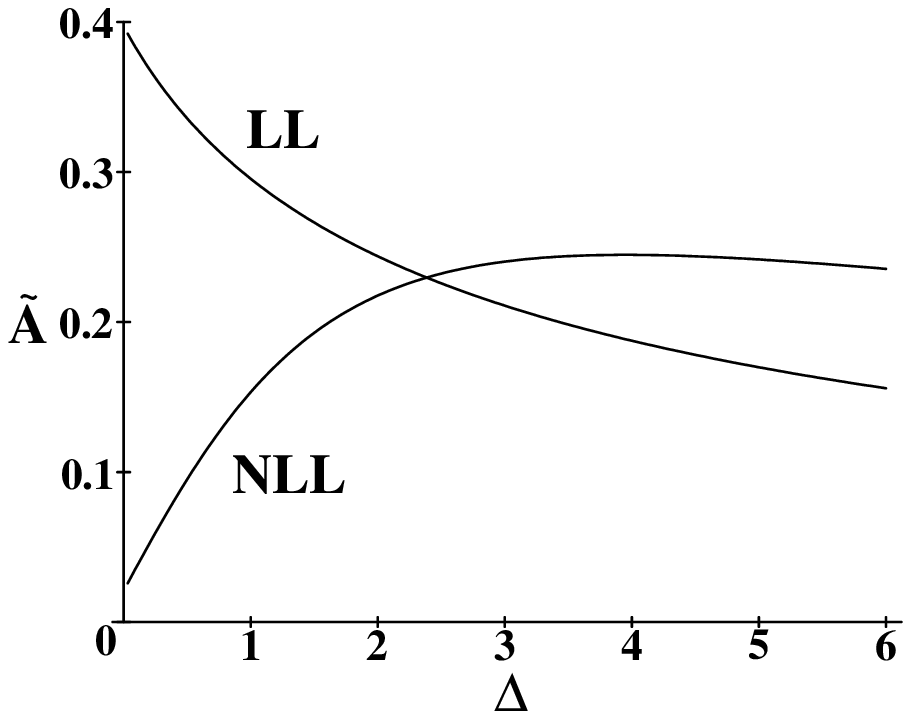}\hskip-0.8in
\epsfxsize 4.0 truein \epsfbox{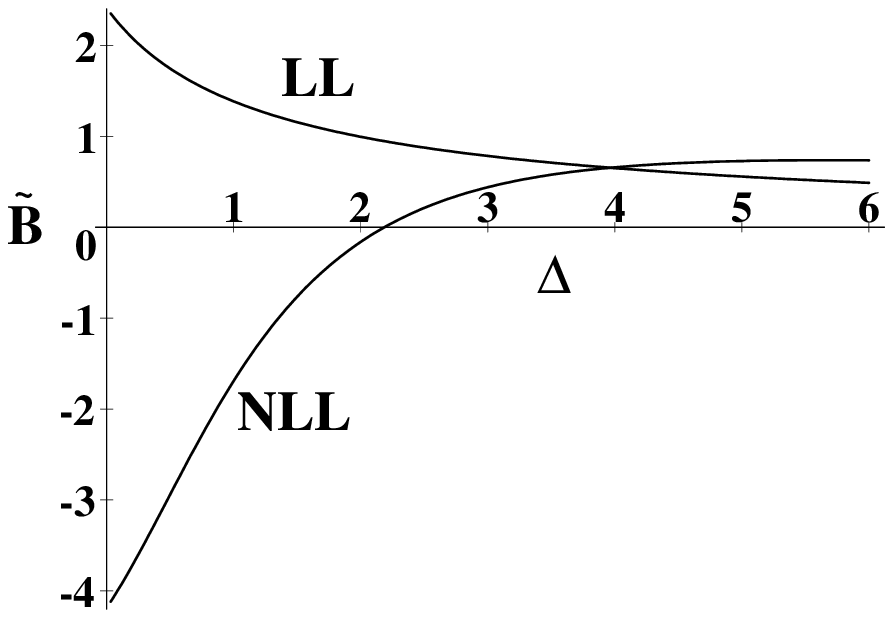}}   
\vskip -.2 cm
\caption[]{
\label{schmidtFigure}
\small Dependence of $\tilde A=\bar\alpha_{s}\chi({1\over2})$ and 
$\tilde B=-{1\over2}\bar\alpha_{s}\chi''({1\over2})$ on $\Delta$ 
for $\alpha_{s}=0.15$, from Ref.~\cite{schmidt}.}
\end{figure}

Fig.~\ref{schmidtFigure} shows the dependence on $\Delta$ of the leading
eigenvalue 
and its second derivative at LL and NLL for $\alpha_{s}=0.15$
in this modified BFKL theory.
Note that the corrections to $\bar\alpha_{s}\chi({1\over2})$ are not 
large for $\Delta\gsim2$ and have weak dependence on $\Delta$ for 
large $\Delta$.  Also, the point $\gamma={1\over2}$ is a maximum for this 
coupling as long as
$\Delta\gsim2.2$.  Thus, the BFKL resummation is 
stable for large enough $\Delta$.  This modification of the BFKL 
resummation predicts a somewhat larger value of the BFKL pomeron 
intercept than the previous proposal.  However,
the implications of a large value of $\Delta$ for the 
phenomenological use of BFKL is open to interpretation.

The third and perhaps most ambitious proposal \cite{salam,CCS} 
to control the large NLL corrections is to systematically 
include the largest collinearly-enhanced, but still 
subleading corrections into LL BFKL.  The most important of these are
energy-scale corrections \cite{cc}.  To understand the origin of these 
corrections, note that in our discussion of BFKL at LL in section 
\ref{bfklatll} we chose to work with the symmetric rapidity 
$\Delta y=\ln\hat s/(p_{a\perp}p_{b\perp})$ as the large  
logarithm to resum.  However, we could equally well have chosen $y^{+}=
\ln x^{+}_{a}/x^{+}_{b}=\ln \hat s/p_{b}^{2}$ 
or $y^{-}=\ln x^{-}_{b}/x^{-}_{a}=
\ln \hat s/p_{a}^{2}$, where $x^{\pm}_{i}$ is the momentum fraction along 
the positive or negative light-cone for the emitted gluon $i$.
These choices are all equivalent at LL because the transverse momenta
are treated as comparable in size; however, at NLL they are inequivalent. 
A change in the logarithm produces a change 
in the NLL kernel and can introduce double transverse logarithms of 
the form $\bar\alpha_{s}\ln^{2}(p_{a\perp}^{2}/p_{b\perp}^{2})$
into the resummation.

Motivated by DGLAP-type resummation \cite{dglap} one finds that the 
appropriate choice is to resum $y^{+}$ when $p_{b\perp}^{2}\gg 
p_{a\perp}^{2}$ and $y^{-}$ when $p_{a\perp}^{2}\gg 
p_{b\perp}^{2}$.  The effect of these changes of the BFKL resummation 
variable was studied in refs.~\cite{cc} and \cite{nll}, and the
corresponding terms in the NLL eigenvalue $\chi^{(2)}(\gamma)$ were 
isolated and resummed in ref.~\cite{salam}.  Additional collinearly-enhanced 
terms due to the effects of the running coupling and the non-singular 
part of the splitting functions have also been considered 
and resummed in Ref.~\cite{CCS}.
A nice discussion of these ideas can be found in ref.~\cite{salamtalk}.

A similar approach, advocated in Refs.~\cite{ABF}, is to use the ``duality''
relations between the BFKL $\ln(1/x)$ and the DGLAP $\ln(Q^2)$ resummations to
incorporate the dominant collinear effects into BFKL at small $x$.  Adding contributions
obtained from the known LO and NLO DGLAP anomalous dimensions, one 
gets a 
``double leading'' expansion for the BFKL function $\chi(\gamma)$, which
is better behaved and more stable in the collinear region near $\gamma=0$.

\begin{figure}[t]	
\centerline{\epsfxsize 4.0 truein \epsfbox{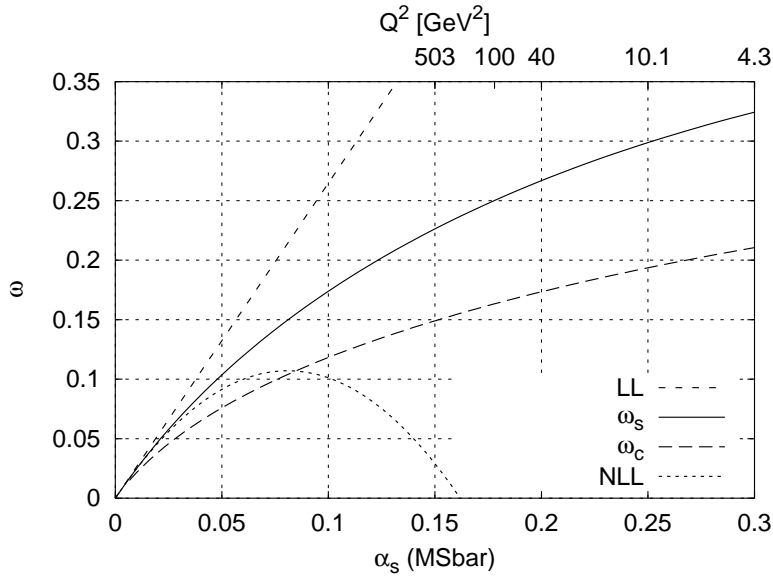}}   
\vskip -.2 cm
\caption[]{
\label{salamFigure}
\small The BFKL Pomeron intercept (here labeled $\omega$) at LL, NLL
and with the collinearly-enhanced resummation included at NLL, from 
Ref.~\cite{CCS}.  The curve $\omega_s$ corresponds to the coefficient
for the exponential rise in $\Delta y$, which is what 
we have focused on in this talk.  The curve $\omega_c$ corresponds to the power growth of the
small-$x$ splitting functions.  Although identical in LL BFKL, these
two coefficients differ at NLL.
}
\end{figure}

Results of the collinearly-enhanced resummation from Ref.\cite{CCS} 
are shown in 
Fig.~\ref{salamFigure}, where the leading eigenvalue 
$\omega_s=\bar\alpha_{s}\chi({1\over2})$ is
 plotted as a function of $\bar\alpha_{s}$.  
As in the other cases the eigenvalue is found to be 
positive after resummation, yielding a value of about $\omega_s=0.27$
for $\alpha_s=0.2$.  In addition the 
characteristic eigenvalue function of $\gamma$ is stable for values of
$\alpha_{s}$ of interest.

The physical implications of the energy-scale dependence
can be seen by further investigating the relation between 
resummation in $y^{\pm}$ and $\Delta y$.  When $p_{b\perp}^{2}\gg
p_{a\perp}^{2}$, the resummation in $y^{+}$ requires the ordering
$x^{+}_{a}>x^{+}_{b}$.  Translating back into the symmetric variable, 
this implies $\Delta y  >\ln(p_{b\perp}/p_{a\perp})$.  Similarly,  
when $p_{a\perp}^{2}\gg 
p_{b\perp}^{2}$, the resummation in $y^{-}$  requires the ordering
$x^{-}_{b}>x^{-}_{a}$, implying
$\Delta y>\ln(p_{a\perp}/p_{b\perp})$.  These constraints hold for 
any two successively emitted gluons.  Therefore, the incorporation
of these collinear effects corresponds to imposing a 
$p_{\perp}$-dependent cut, $y_{i+1}-y_{i}>|\ln(p_{i\perp}/p_{i+1\perp})|$, 
on the separation in rapidity between the 
neighboring gluons.  Since this cut is very similar to the rapidity 
veto of proposal two, it is understandable that when both the 
collinear resummation and the rapidity veto are included, as studied 
in Ref.~\cite{frv}, the dependence on the parameter $\Delta$ was 
significantly reduced, even for small $\Delta$.

\section{Phenomenology of BFKL}

Although the theoretical studies of BFKL have been focused on its 
behavior at NLL, the level of most phenomenological studies is still
at LL.  To treat a process consistently at NLL, one also must 
incorporate the ${\cal O}(\alpha_{s})$ corrections to the impact 
factors.  Although these corrections are known for some processes, 
at least at the amplitude level, they have yet to be incorporated into 
a consistent calculation suitable for phenomenological studies.

In this section I will discuss several probes of BFKL physics in
hadron-hadron, lepton-hadron, and $\gamma^{\ast}\gamma^{\ast}$ 
collisions.  I will only consider processes for which the relevant
transverse scales $p_{a\perp}, p_{b\perp}$ on both sides of the BFKL
ladder can be considered perturbative.  In particular, I will not
consider the inclusive $F_2(x,Q^2)$ in DIS, because it is necessarily
dependent on non-perturbative inputs.

\subsection{BFKL probes at the Tevatron}

One of the most thoroughly studied searches for BFKL physics has been by the 
D\O\ collaboration in $p\bar p$ collisions 
at the Fermilab Tevatron.  As in all BFKL experimental studies, the
 basic idea is to analyze the 
events in a configuration which most closely approximates that used 
in the BFKL resummation.  Jets, with transverse momentum above some 
$E_{\perp{\rm min}}$, are tagged and ordered in rapidity.  Then one 
defines observables as a function of $\Delta y=y_{a}-y_{b}$, 
where $y_{a}$ and $y_{b}$ are the rapidities of the 
most forward and backward jets, respectively.  

The most natural BFKL signal would be the power-law growth in the partonic 
cross section with $\Delta y$, as in Eq.~(\ref{saddle}).  However, at 
fixed center-of-mass energy this growth is swamped by the 
effects of steeply falling parton distribution functions (PDFs) which 
are relevant when far forward or backward jets are produced.  Thus, the 
first observable 
to be considered was the decorrelation in azimuthal angle 
between the two tagged jets as a function of $\Delta y$.  Physically, 
this effect is easy to understand.  In the Born approximation, only 
two jets are produced, and by momentum conservation they must be back-to-back. 
However, as the rapidity interval increases, there is more room for 
additional jets, and the tagged jets become decorrelated.  
This can be seen directly in the BFKL solution (\ref{sold}).  At 
small $\Delta y$, all terms in the Fourier series in 
$\tilde\phi=\phi_{a}-\phi_{b}-\pi$ are approximately equal, producing 
a delta-function in $\tilde\phi$ which forces the two jets to be back 
to back.  As $\Delta y$ increases, the higher order terms become 
smaller and smaller compared to the leading $n=0$ term, so the 
jets become completely decorrelated \cite{vddi}.  The simplest 
observable to display this effect is the moment 
$\langle\cos\tilde\phi\rangle$, which goes to 1 if the jet azimuthal 
angles are completely correlated and goes to 0 if they are completely 
decorrelated \cite{stirling}.

Even before the D\O\ analysis was completed, however, it was realized 
that there was a serious problem in using LL BFKL for phenomenological 
analyses at hadron colliders: the BFKL resummation includes 
the contribution of energetically disfavored or 
disallowed configurations in its predictions.  In principle these 
configurations are 
subleading, but in practice they are very important \cite{vddii}.  (In fact 
these effects could be considered to be a foreshadowing of the large 
corrections to BFKL at NLL.)
We can understand this effect by considering the 
Feynman $x$-values used in the
PDFs.  The exact values are given by conservation of light-cone
momentum along the beam axis and can be written
\begin{eqnarray}
    x_{a}&=&\frac{1}{\sqrt{s}}\left(p_{a\perp}e^{y_{a}}+
    p_{b\perp}e^{y_{b}}+\sum_{i}
    k_{i\perp}e^{y_{i}}\right)\nonumber\\
    x_{b}&=&\frac{1}{\sqrt{s}}\left(p_{a\perp}e^{-y_{a}}+
    p_{b\perp}e^{-y_{b}}+\sum_{i}
    k_{i\perp}e^{-y_{i}}\right)\ ,
    \label{pdfexact}
\end{eqnarray}
where the sum is over all partons produced in the event.
In the ``naive'' LL BFKL one only keeps the leading contributions,
\begin{eqnarray}
    x^{0}_{a}&=&\frac{p_{a\perp}e^{y_{a}}}{\sqrt{s}}\nonumber\\
    x^{0}_{b}&=&\frac{p_{b\perp}e^{-y_{b}}}{\sqrt{s}}\ .
    \label{pdfnaive}
\end{eqnarray}
That is, one convolutes the analytic LL BFKL solution (\ref{sold}) with 
the impact factors, using the PDFs evaluated at $x^{0}_{a,b}$.
However, the true $x_{a,b}$ and are always larger than $x^{0}_{a,b}$, 
and the energy-momentum constraints $x_{a,b}<1$ are not enforced in the naive
BFKL calculation.  If the PDFs vary strongly with 
$x$, this can greatly overestimate the contributions from multi-jet
events.

With the analytic LL BFKL solution (\ref{sold}) 
the phase space of the intermediate gluons has already been integrated over, 
so there is no choice but to use the leading $x^{0}$'s (\ref{pdfnaive}) 
in the PDFs.  However, in a BFKL Monte Carlo 
solution \cite{bfklmc,orrstir} one generates the gluon ladder directly as in 
Eq.~(\ref{llbfkl}).  Thus, one has information on all the produced 
partons, and one can enforce energy conservation on the solution by 
using the exact $x$'s (\ref{pdfexact}) in the PDFs.  This greatly improves 
the reliability of the BFKL prediction.

\begin{figure}[t]	
\centerline{\epsfxsize 3.0 truein \epsfbox{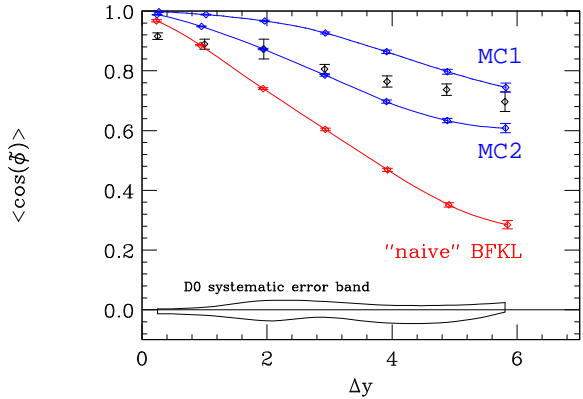}}
\vskip 0.0 cm
\caption[]{
\label{cosphi}
\small 
Azimuthal angle decorrelation as a function of rapidity interval. 
The D\O\ data is from \cite{decor}, the lowest curve is ``naive'' BFKL, 
the two upper curves are BFKL Monte Carlo predictions. The 
correlated systematic error is due to jet energy scale uncertainty.}
\end{figure}

In Fig.~\ref{cosphi} we show the D\O\ azimuthal decorrelation
data from Ref.~\cite{decor} compared with
the naive LL BFKL and a Monte Carlo BFKL calculation with energy
conservation included.  The data shows 
$\langle\cos\tilde\phi\rangle$ as a function of $\Delta y$ at 
$\sqrt{s}=1.8$ TeV, where the
tagging jets are required to have $E_{\perp}>20$ GeV.  The naive BFKL
severely overestimates the rate of decorrelation with $\Delta y$, due 
to the overweighting of energetically disfavored or disallowed 
configurations.   The two BFKL Monte Carlo curves are produced by the
Monte Carlo in Ref.~\cite{bfklmc}.  Both  use the exact $x$'s in the
PDFs, but use different approximations to relate the partonic 
to the hadronic cross sections in the high energy limit.  
The difference
between the two approximations is subleading in the high energy limit.
The BFKL Monte Carlo with energy conservation certainly works much better
than the naive BFKL, but the subleading uncertainties are still 
sizeable, as displayed by the difference between the two Monte Carlo
predictions.

An argument against the decorrelation measurement as a signal of BFKL
is that it probes
the region around $\tilde\phi\approx0$, which is also most sensitive to
Sudakov logarithms. It would be nicer to probe directly the rise
in partonic cross section with the partonic energy as in Eq.~(\ref{saddle}).
This became possible when the Tevatron collider was run at the
lower energy of 630 GeV, 
allowing a comparison of dijet production at two different center-of-mass
energies.  By binning the events in the partonic $x_a, x_b$ values rather 
than in $\Delta y$, the dependence on the PDFs should cancel in the ratio
\begin{equation}
R\ =\ \frac{\sigma(\sqrt{s_1}=1800 \mbox{\rm GeV})}{\sigma(
\sqrt{s_2}=630 \mbox{\rm GeV})}\ .
\end{equation}
This is the original proposal of Mueller and Navelet \cite{mn}.  Using 
the asymptotic saddle point approximation, one obtains a prediction of
\begin{equation}
R_{\mbox{\rm BFKL}}\ =\ \frac{e^{A(\Delta y_1-\Delta y_2)}}{\sqrt{\Delta y_1/
\Delta y_2}}\ ,
\label{bfklratio}
\end{equation}
where $\Delta y_i\approx\ln(\hat s_i/p_{a\perp}p_{b\perp})
\approx\ln(\hat s_i/E_{\perp{\rm min}}^2)$.  This ratio was measured by
the D\O\ collaboration, in several $x$ bins, with the following cuts:
$E_{\perp\mbox{\rm min}}>20$ GeV, $|y|<3$, $\Delta y>2$, and 
$400<Q^2=E_{a\perp}E_{b\perp}<1000$ GeV$^2$.
Using the formula (\ref{bfklratio}), 
a BFKL Pomeron intercept of $A=0.65\pm0.07$ was extracted \cite{ratio}.

This measurement is noteworthy in that it is probably the only current
measurement
which shows a rise in the cross section that is {\em larger} than the LL BFKL
prediction.  Using $\alpha_s(20\mbox{ GeV})=0.17$ in Eq.~(\ref{Avalue}),
one obtains a LL prediction of $A=0.45$, which is almost 3 standard deviations
below the extracted value.  However, as discussed in Ref.~\cite{jeppe}, one
must be careful in interpreting this measurement.  First, the extraction
of the BFKL Pomeron intercept from (\ref{bfklratio}) assumes that the 
asymptotic BFKL expression is valid, and the experimental cuts 
and precise definition of the $x$'s do not significantly affect the 
asymptotics.  In particular the cut on $Q^2$ was seen to slow the approach
to asymptotics, resulting in a smaller predicted value for the ratio $R$.
The inclusion of energy conservation
via a BFKL Monte Carlo further reduced the predicted ratio.  
Finally, it was shown that the use of equal $E_\perp$ cuts on both of the
tagging jets introduces the same large Sudakov logarithms that plague the
decorrelation measurement. Thus, it seems
unlikely that the large ratio found by the D\O\ collaboration can be
attributed to perturbative BFKL.

\subsection{BFKL probes at HERA}

It is also possible to look for the BFKL rise in the cross section in
deep inelastic scattering (DIS) by tagging a forward jet 
\cite{muellerdis,otherdis}.
Referring to the high energy factorization picture of Fig.~\ref{factorize},
the DIS setup consists of the scattering of an off-shell photon with
virtuality $Q^2$ and Bjorken $x_{bj}=Q^2/s_{ep}$ on the left, the scattering
of a forward jet of momentum fraction $x_{\rm jet}=E_{\rm jet}/E_p$ and 
transverse
momentum $p_\perp$ on the right, connected by the BFKL ladder of gluon
emissions in the
middle.  If $Q^2\sim p_\perp^2\gg\Lambda_{QCD}^2$, then the BFKL evolution
is perturbative.  The large logarithm that is resummed is 
$\ln(x_{\rm jet}/x_{bj})$.

This DIS setup has several advantages over the $p\bar p$ setup, due to its 
asymmetric nature.  Note that in the high energy limit, the PDF of the
proton is evaluated at $x_{\rm jet}$.  Thus, with $x_{\rm jet}$ fixed,
one can vary the rapidity
interval at a single collider energy by varying $x_{bj}$, without any
change in the PDF. In addition, this suggests that the energy conservation
effects mentioned above may be less important, or at least not so strongly
dependent on $x_{bj}$.  Finally, the resummation in 
$\ln(x_{\rm jet}/x_{bj})$, rather than in the jet rapidity intervals, is
more natural here, and perhaps more stable theoretically, since it is
corresponds to the standard DGLAP evolution variable when $Q^2>p_{\perp}^2$.
The only major disadvantage is that the virtual photon impact factor is
more complicated theoretically than the gluon or quark impact factors.
Indeed, it has not been calculated completely at NLO.

\begin{figure}[t]	
\center{\epsfig{file=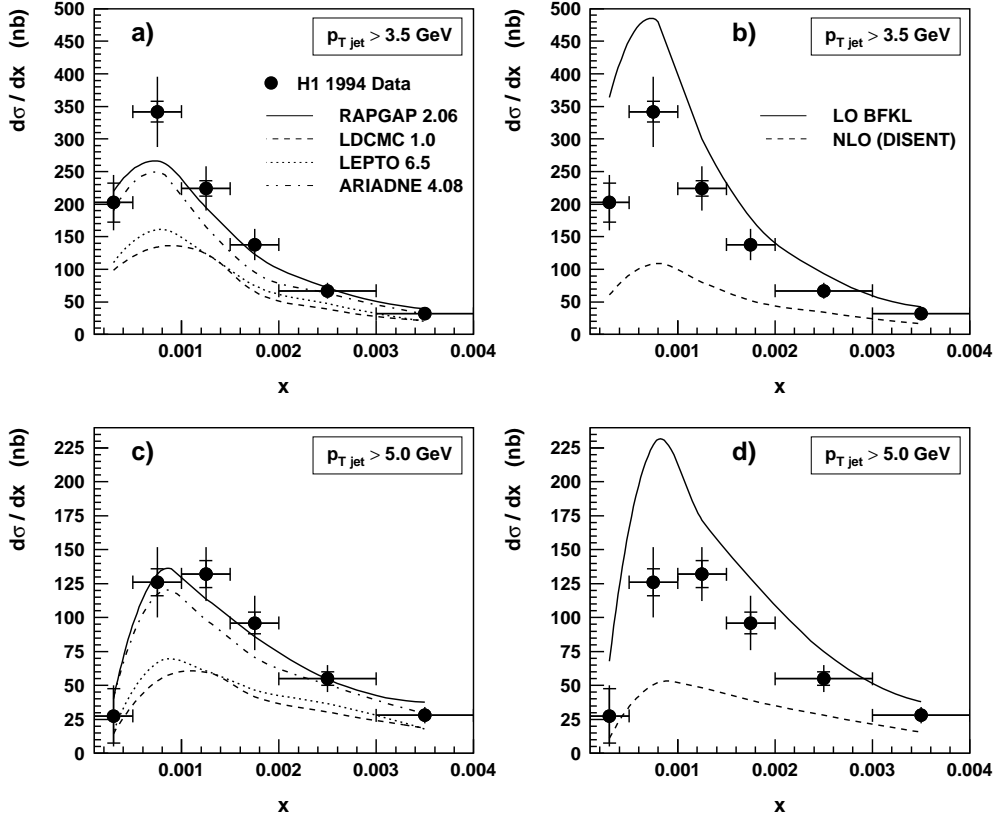,width=120mm,bbllx=50pt,bblly=180pt,bburx=520pt,bbury=650pt}}
\vskip -0.5 cm
\caption[]{
\label{honedata}
\small 
Forward jet cross-sections as a function of Bjorken-$x$ from H1 \cite{Hone}
for two $p_{\perp}$ cuts of 3.5 and 5 GeV.  The data in (a) and (c) are 
compared with Monte-Carlo model predictions, including hadronization.  The
data in (b) and (d) are compared with partonic NLO ${\cal O}(\alpha_s^2)$ 
and LL BFKL calculations.}
\end{figure}

\begin{figure}[t]	
\centerline{\epsfxsize 3.5 truein \epsfbox{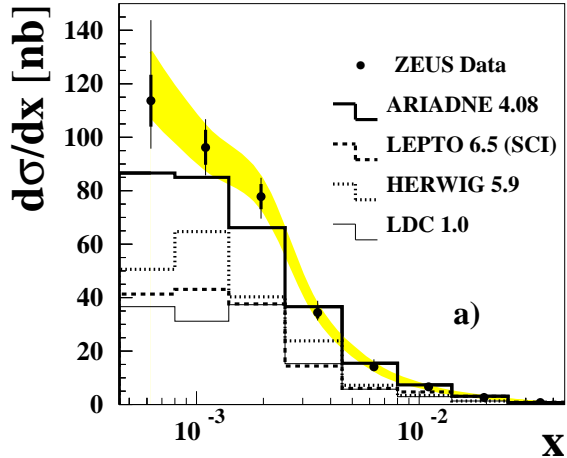}}
\vskip -1.0 cm
\caption[]{
\label{zeusdata}
\small 
Forward jet cross-sections as a function of Bjorken-$x$ from ZEUS \cite{zeus}
for $p_{\perp}>5$ GeV.  The data are compared with several 
Monte-Carlo model predictions at the hadron level.}
\end{figure}

Both the H1 \cite{Hone} and ZEUS \cite{zeus} collaborations have measured
this forward jet cross section.  
The main experimental cuts on the forward jet
itself are $x_{\rm jet}>0.035$, $E_{\perp{\rm jet}}>3.5$ and 5 GeV for H1,
$x_{\rm jet}>0.036$, $E_{\perp{\rm jet}}>5$ GeV for ZEUS, and
$0.5<E_{\perp{\rm jet}}^2/Q^2<2$ for both. 
%
The H1 data 
is displayed in Fig.~\ref{honedata}.
In Figs.~\ref{honedata}(a) and
(c) the data is compared against several hadron-level Monte Carlo shower 
models attached to lowest order QCD matrix elements.  The two that 
agree the best with the data are ARIADNE \cite{ariadne}
and RAPGAP \cite{rapgap}.  ARIADNE is based on the colour dipole model
for gluon radiation which, like BFKL, lacks ordering in $k_\perp$.  Gluon
radiation in RAPGAP is based on DGLAP evolution, but this model also includes
a resolved photon contribution to the basic QCD production mechanism.
In Figs.~\ref{honedata}(b) and
(d) the H1 data is compared against a LL BFKL prediction \cite{bartels}
and a NLO QCD ${\cal O}(\alpha_s^2)$ prediction \cite{disent}, both at the
parton levels.  The NLO calculation significantly underestimates the data 
at small $x_{bj}$, whereas the BFKL calculation overestimates it.  
This is not unreasonable, given that NLL corrections to BFKL are expected 
to reduce the rise at small $x_{bj}$, and that the kinematic cuts could not
be included exactly in the calculation.  
%
The ZEUS data \cite{zeus}, shown in Fig.~\ref{zeusdata}, similarly is 
far above a NLO QCD calculation
\cite{mirkes}, but below the LL BFKL expectations.
A fit \cite{Contreras} 
to the data from both experiments using the LL BFKL cross section 
yielded an effective BFKL pomeron intercept corresponding to $A=0.43\pm0.025
\mbox{\rm (stat)}\pm 0.025\mbox{\rm (sys)}$, compared to the LL prediction
from Eq.~(\ref{Avalue}) of $A=0.75$ for $\alpha_s=0.28$ at $Q^2=10$ GeV$^2$.

Recently, several forward jet
calculations with different approaches have shown good
agreement with the data.  One calculation \cite{kmo} is based on LL
BFKL, but 
modified by a consistency condition containing effects similar to the 
dominant NLL energy-scale effects discussed in section \ref{sec:nllbig}.
A second calculation \cite{salamjung} is with a hadron-level Monte Carlo 
generator, based on the CCFM evolution
\cite{ccfm}, which is designed to agree with both DGLAP and BFKL in their 
respective regimes of reliability.  
Since both of these calculations can be considered LL BFKL, with some
dominant subleading corrections included, this looks promising for BFKL.
A third approach \cite{potter} which also fits the data is a NLO calculation 
that includes a resolved photon contribution, similar in spirit to the 
RAPGAP Monte Carlo.  Interestingly, it appears that the success of this
approach relies not on the evolution in $Q^2$ allowed by the inclusion of
the photon PDF, but in the fact that it effectively approximates one term
higher in $\alpha_s$, via the NLO resolved piece.
This is not incompatible with BFKL, since the new ${\cal O}(\alpha_s^3)$
contribution also includes the first gluon emission in the BFKL ladder.  
It is an interesting question to ask how the approximations in this
picture of forward jet production mesh with those in the BFKL picture.

\subsection{BFKL probes in $\gamma^{\ast}\gamma^{\ast}$}

Another standard BFKL measurement \cite{bhs} is to observe the
total $\gamma^{\ast}\gamma^{\ast}$ hadronic cross section as a function
of $\sqrt{s_{\gamma\gamma}}$.  This can be extracted from
the process $e^+e^-\rightarrow e^+e^-+$ hadrons by tagging on the forward
and backward electrons.  In this case the two independent scatterings
of Fig.~\ref{factorize} are the off-shell photons of momenta $Q_1^2$ and 
$Q_2^2$ that break up into color dipoles ($q\bar q$ pairs at leading order), 
which are then connected by the BFKL ladder. The large logarithm here is
$Y\approx\ln(s_{\gamma\gamma}/\sqrt{Q_1^2Q_2^2})$, in direct analogy to
the hadron-hadron case.

This experiment has the advantage that, for large enough $Q^2_{1,2}$,
there are no nonperturbative PDF inputs, so that in principle it is the
cleanest probe, theoretically.  However,
from a purely calculational point of view, this process may be more 
complicated since it involves two off-shell photon impact factors.
In particular, the cross-channel gluon on which the BFKL ladder is built
does not even appear until NNLO in a standard perturbative QCD calculation..  

\begin{figure}[t]	
\centerline{\epsfxsize 3.0 truein \epsfbox{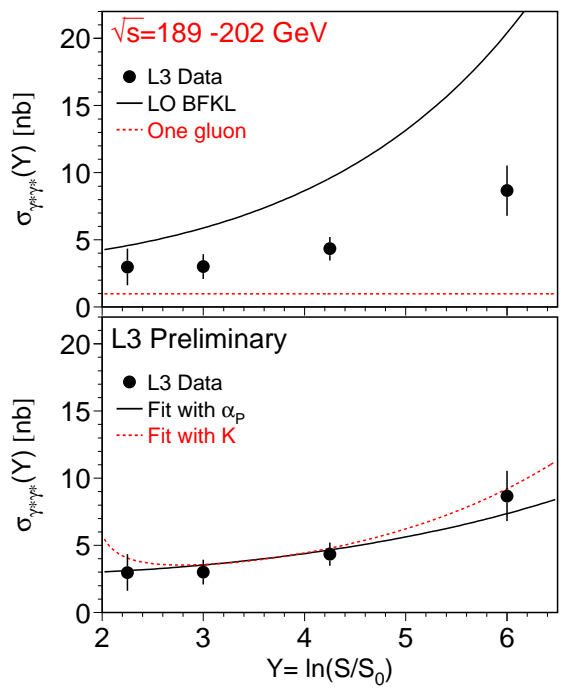}}
\vskip 0.0 cm
\caption[]{
\label{lthree}
\small 
Two-photon cross-sections, $\sigma_{\gamma^{\ast}\gamma^{\ast}}$,
after subtraction of the LO $\gamma^{\ast}\gamma^{\ast}
\rightarrow q\bar q$ contribution at $\sqrt{s}\simeq189-202$ GeV
and $4<Q^2_{1,2}<40$ GeV$^2$, 
from Ref.~\cite{lthreeii}.
The top figure compares the data with LO BFKL in the saddle point 
approximation with $A=\alpha_P-1=0.53$ and with the one-gluon exchange
diagram.  The lower figure shows fits of the saddle-point BFKL 
expression to the data, allowing either $\alpha_P$ or the normalization $K$ to
vary. }
\end{figure}

The $\gamma^{\ast}\gamma^{\ast}$ hadronic cross section
has been measured by the L3 \cite{lthreei,lthreeii} and the OPAL \cite{opal}
collaborations at LEP.  In Fig.~\ref{lthree} we display the most recent
preliminary L3 data \cite{lthreeii}.  The data clearly shows a rise
with $Y$ as expected by BFKL, but much less steep than LO BFKL.  However,
the data is above both LO and NLO QCD predictions \cite{cdft}.
In addition the data is also above the prediction from the single-gluon
exchange contribution (evaluated using the average value
of $\langle Q^2_{1,2}\rangle=15$ GeV$^2$).  In the lower figure the
data is fitted to the asymptotic BFKL prediction
\begin{equation}
\sigma_{\gamma^{\ast}\gamma^{\ast}}\ =\ \frac{\sigma_0}{\sqrt{Q_1^2Q^2_2Y}}
e^{AY}\ ,
\end{equation}
with either the overall normalization or the BFKL intercept $A=\alpha_P-1$
left as a free parameter.  The preliminary result in the latter
case gives $A=0.36\pm0.02$, which is more in line with NLO expectations.
The OPAL preliminary measurements \cite{opal} give qualtitatively similar
results, but with less statistical significance.

\section{Summary and Conclusions.}

In this talk I have presented some recent results in BFKL physics, 
both in the theory and in its phenomenological applications.  On the
theoretical side, the focus has been on the large NLL corrections.  
At this time, it seems fairly safe to say that the original catastrophe
of falling, or even negative, cross sections has been averted.  By 
understanding the origin of these effects in the collinear behavior of
the gluons, one can reorganize the resummation in order to move the
dominant corrections back into the LL theory.  Then the NLL prediction
for the BFKL intercept is stable and slightly smaller than the standard
LL prediction.

On the phenomenological side, the results look suggestive, especially
in the DIS and $\gamma^{\ast}\gamma^{\ast}$ data.  The cross sections
are significantly above the state-of-the-art NLO QCD calculations, as
they should be for a BFKL-enhanced observable.  They are also in the
range one might expect from a NLL BFKL calculation.  
However, the lesson learned from the analysis of the hadron-hadron 
experiments 
is that one must be very careful to consider how experimental cuts
and kinematic effects, such as energy conservation, will affect these
predictions.  Although the asymmetric configuration of the DIS forward
jet experiments and the lack of nonperturbative PDF-dependence
in the $\gamma^{\ast}\gamma^{\ast}$ (assuming $Q^2$ is large enough) may
make these observables less susceptible to large subleading effects, 
a thorough phenomenological analysis is certainly warranted.

With the completion and understanding of the NLL corrections to the BFKL
kernel in hand, the next phase is to bring the phenomenological analyses up to
the same NLL level.  So far the major emphasis has been on the NLL
BFKL intercept, but to make a full NLL prediction, with a reliable
normalization, one also needs to combine this with NLO impact factors.
In particular the NLO impact factor for the off-shell photon is 
crucial for the both the DIS and $\gamma^{\ast}\gamma^{\ast}$ analyses.
In addition the analyses must be performed in such a way as to treat the
kinematics and cuts as accurately as possible.  Promising steps
in this direction are Refs.~\cite{kmo,salam}, which incorporate the 
largest NLL corrections into BFKL via a consistency condition, or via
a CCFM Monte Carlo, respectively.  
Another useful exercise would be to incorporate
the largest NLL corrections to the BFKL ladder into the BFKL Monte Carlos
\cite{bfklmc,orrstir} and to modify them for use in the other
experimental environments.  This would be helpful for gauging the sensitivity
to subleading kinematic effects (which arise at least as much  
from the impact factors as from the actual BFKL ladder).
However, the greatest progress would come with a full NLL 
calculation with full NLO impact
factors included.  There is much work left to be done.


\begin{thebibliography}{99}

\bibitem{FKL}
E.~A.~Kuraev, L.~N.~Lipatov and V.~S.~Fadin,
Sov.\ Phys.\ JETP {\bf 44}, 443 (1976)
[Zh.\ Eksp.\ Teor.\ Fiz.\  {\bf 71}, 840 (1976)];\\
E.~A.~Kuraev, L.~N.~Lipatov and V.~S.~Fadin,
Sov.\ Phys.\ JETP {\bf 45}, 199 (1977)
[Zh.\ Eksp.\ Teor.\ Fiz.\  {\bf 72}, 377 (1977)];\\
I.~I.~Balitsky and L.~N.~Lipatov,
Sov.\ J.\ Nucl.\ Phys.\  {\bf 28}, 822 (1978)
[Yad.\ Fiz.\  {\bf 28}, 1597 (1978)].

\bibitem{schmidt}
C.~R.~Schmidt,
Phys.\ Rev.\ D {\bf 60}, 074003 (1999)
[hep-ph/9901397].

\bibitem{nll}
V.~S.~Fadin and L.~N.~Lipatov,
Phys.\ Lett.\ B {\bf 429}, 127 (1998)
[hep-ph/9802290].

\bibitem{ddsii}
V.~Del Duca and C.~R.~Schmidt,
Phys.\ Rev.\ D {\bf 59}, 074004 (1999)
[hep-ph/9810215].

\bibitem{cc}
G.~Camici and M.~Ciafaloni,
Phys.\ Lett.\ B {\bf 412}, 396 (1997)
[Erratum-ibid.\ B {\bf 417}, 390 (1997)]
[hep-ph/9707390];\\
Phys.\ Lett.\ B {\bf 430}, 349 (1998)
[hep-ph/9803389].

\bibitem{Kovchegov}
Y.~V.~Kovchegov and A.~H.~Mueller,
Phys.\ Lett.\ B {\bf 439}, 428 (1998)
[hep-ph/9805208].

\bibitem{Armesto}
N.~Armesto, J.~Bartels and M.~A.~Braun,
Phys.\ Lett.\ B {\bf 442}, 459 (1998)
[hep-ph/9808340].

\bibitem{levin}
E.~Levin,
hep-ph/9806228.

\bibitem{CCS}
M.~Ciafaloni and D.~Colferai,
Phys.\ Lett.\ B {\bf 452}, 372 (1999)
[hep-ph/9812366];\\
M.~Ciafaloni, D.~Colferai and G.~P.~Salam,
Phys.\ Rev.\ D {\bf 60}, 114036 (1999)
[hep-ph/9905566].


\bibitem{ross}
D.~A.~Ross,
Phys.\ Lett.\ B {\bf 431}, 161 (1998)
[hep-ph/9804332].

\bibitem{ball}
R.~D.~Ball and S.~Forte,
hep-ph/9805315.

\bibitem{blumlein}
J.~Blumlein and A.~Vogt,
Phys.\ Rev.\ D {\bf 58}, 014020 (1998)
[hep-ph/9712546];\\
J.~Blumlein, V.~Ravindran, W.~L.~van Neerven and A.~Vogt,
hep-ph/9806368.

\bibitem{blm}
S.~J.~Brodsky, V.~S.~Fadin, V.~T.~Kim, L.~N.~Lipatov and G.~B.~Pivovarov,
JETP Lett.\  {\bf 70}, 155 (1999)
[hep-ph/9901229].

\bibitem{blmone}
S.~J.~Brodsky, G.~P.~Lepage and P.~B.~Mackenzie,
Phys.\ Rev.\ D {\bf 28}, 228 (1983).

\bibitem{bo}
B.~Andersson, G.~Gustafson and J.~Samuelsson,
Nucl.\ Phys.\ B {\bf 467}, 443 (1996).

\bibitem{liptalk} L.N.~Lipatov, talk presented
at the $4^{\rm th}$ Workshop on Small-$x$ and Diffractive Physics,
Fermi National Accelerator Laboratory, Sept. 17-20, 1998.

\bibitem{salam}
G.~P.~Salam,
JHEP {\bf 9807}, 019 (1998)
[hep-ph/9806482].

\bibitem{dglap}
Y.~L.~Dokshitzer,
Sov.\ Phys.\ JETP {\bf 46}, 641 (1977)
[Zh.\ Eksp.\ Teor.\ Fiz.\  {\bf 73}, 1216 (1977)];\\
V.~N.~Gribov and L.~N.~Lipatov,
Yad.\ Fiz.\  {\bf 15}, 781 (1972)
[Sov.\ J.\ Nucl.\ Phys.\  {\bf 15}, 438 (1972)];\\
G.~Altarelli and G.~Parisi,
Nucl.\ Phys.\ B {\bf 126}, 298 (1977).

\bibitem{salamtalk}
G.~P.~Salam,
Acta Phys.\ Polon.\ B {\bf 30}, 3679 (1999)
[hep-ph/9910492].


\bibitem{ABF}
G.~Altarelli, R.~D.~Ball and S.~Forte,
Nucl.\ Phys.\ B {\bf 575}, 313 (2000)
[hep-ph/9911273];\\
G.~Altarelli, R.~D.~Ball and S.~Forte,
Nucl.\ Phys.\ B {\bf 599}, 383 (2001)
[hep-ph/0011270].

\bibitem{frv}
J.~R.~Forshaw, D.~A.~Ross and A.~Sabio Vera,
Phys.\ Lett.\ B {\bf 455}, 273 (1999)
[hep-ph/9903390].

\bibitem{vddi}
V.~Del Duca and C.~R.~Schmidt,
Phys.\ Rev.\ D {\bf 49}, 4510 (1994)
[hep-ph/9311290].

\bibitem{stirling}
W.~J.~Stirling,
Nucl.\ Phys.\ B {\bf 423}, 56 (1994)
[hep-ph/9401266].

\bibitem{vddii}
V.~Del Duca and C.~R.~Schmidt,
Phys.\ Rev.\ D {\bf 51}, 2150 (1995)
[hep-ph/9407359].

\bibitem{bfklmc}
C.~R.~Schmidt,
Phys.\ Rev.\ Lett.\  {\bf 78}, 4531 (1997)
[hep-ph/9612454].

\bibitem{orrstir}
L.~H.~Orr and W.~J.~Stirling,
Phys.\ Rev.\ D {\bf 56}, 5875 (1997)
[hep-ph/9706529].

\bibitem{decor}
B.~Abbott {\it et al.}  [D0 Collaboration],
FERMILAB-CONF-97-371-E
{\it Contributed to 18th International Symposium on Lepton - Photon Interactions (LP 97), Hamburg, Germany, 28 Jul - 1 Aug 1997, and  Contributed
to International Europhysics Conference on High-Energy Physics (HEP 97), Jerusalem, Israel, 19-26 Aug 1997};\\
S.~Abachi {\it et al.}  [D0 Collaboration],
Phys.\ Rev.\ Lett.\  {\bf 77}, 595 (1996)
[hep-ex/9603010].

\bibitem{mn}
A.~H.~Mueller and H.~Navelet,
Nucl.\ Phys.\ B {\bf 282}, 727 (1987).

\bibitem{ratio}
B.~Abbott {\it et al.}  [D0 Collaboration],
Phys.\ Rev.\ Lett.\  {\bf 84}, 5722 (2000)
[hep-ex/9912032].

\bibitem{jeppe}
J.~R.~Andersen, V.~Del Duca, S.~Frixione, C.~R.~Schmidt and W.~J.~Stirling,
JHEP {\bf 0102}, 007 (2001)
[hep-ph/0101180].

\bibitem{muellerdis}
A.~Mueller,
Nucl.\ Phys.\ B (Proc.\ Suppl.) {\bf 18}C, 125 (1991).

\bibitem{otherdis}
J.~Kwiecinski, A.~D.~Martin and P.~J.~Sutton,
Phys.\ Rev.\ D {\bf 46}, 921 (1992);\\
J.~Bartels, A.~de Roeck and M.~Loewe,
Z.\ Phys.\ C {\bf 54}, 635 (1992);\\
W.~Tang,
Phys.\ Lett.\ B {\bf 278}, 363 (1992).

\bibitem{Hone}
C.~Adloff {\it et al.}  [H1 Collaboration],
Nucl.\ Phys.\ B {\bf 538}, 3 (1999)
[hep-ex/9809028].

\bibitem{zeus}
J.~Breitweg {\it et al.}  [ZEUS Collaboration],
Eur.\ Phys.\ J.\ C {\bf 6}, 239 (1999)
[hep-ex/9805016].
\bibitem{ariadne}
L.~Lonnblad,
Comput.\ Phys.\ Commun.\  {\bf 71}, 15 (1992).

\bibitem{rapgap}
H.~Jung,
Comput.\ Phys.\ Commun.\  {\bf 86}, 147 (1995).
H.~Jung, L.~Jonsson and H.~Kuster,
Eur.\ Phys.\ J.\ C {\bf 9}, 383 (1999)
[hep-ph/9903306].

\bibitem{bartels}
J.~Bartels, V.~Del Duca, A.~De Roeck, D.~Graudenz and M.~Wusthoff,
Phys.\ Lett.\ B {\bf 384}, 300 (1996)
[hep-ph/9604272].

\bibitem{disent}
S.~Catani and M.~H.~Seymour,
Phys.\ Lett.\ B {\bf 378}, 287 (1996)
[hep-ph/9602277].
S.~Catani and M.~H.~Seymour,
Nucl.\ Phys.\ B {\bf 485}, 291 (1997)
[Erratum-ibid.\ B {\bf 510}, 503 (1997)]
[hep-ph/9605323].

\bibitem{mirkes}
E.~Mirkes and D.~Zeppenfeld,
Phys.\ Rev.\ Lett.\  {\bf 78}, 428 (1997)
[hep-ph/9609231].

\bibitem{Contreras}
J.~G.~Contreras, R.~Peschanski and C.~Royon,
Phys.\ Rev.\ D {\bf 62}, 034006 (2000)
[hep-ph/0002057];\\
J.~G.~Contreras,
Phys.\ Lett.\ B {\bf 446}, 158 (1999)
[hep-ph/9812255].


\bibitem{kmo}
J.~Kwiecinski, A.~D.~Martin and J.~J.~Outhwaite,
Eur.\ Phys.\ J.\ C {\bf 9}, 611 (1999)
[hep-ph/9903439].
 
\bibitem{salamjung}
H.~Jung and G.~P.~Salam,
Eur.\ Phys.\ J.\ C {\bf 19}, 351 (2001)
[hep-ph/0012143].

\bibitem{ccfm}
M.~Ciafaloni,
Nucl.\ Phys.\ B {\bf 296}, 49 (1988);\\
S.~Catani, F.~Fiorani and G.~Marchesini,
Phys.\ Lett.\ B {\bf 234}, 339 (1990);\\
S.~Catani, F.~Fiorani and G.~Marchesini,
Nucl.\ Phys.\ B {\bf 336}, 18 (1990).

\bibitem{potter}
G.~Kramer and B.~Potter,
Phys.\ Lett.\ B {\bf 453}, 295 (1999)
[hep-ph/9901314].

\bibitem{bhs}
J.~Bartels, A.~De Roeck and H.~Lotter,
Phys.\ Lett.\ B {\bf 389}, 742 (1996)
[hep-ph/9608401];\\
S.~J.~Brodsky, F.~Hautmann and D.~E.~Soper,
Phys.\ Rev.\ D {\bf 56}, 6957 (1997)
[hep-ph/9706427].

\bibitem{lthreei}
M.~Acciarri {\it et al.}  [L3 Collaboration],
Phys.\ Lett.\ B {\bf 453}, 333 (1999).

\bibitem{lthreeii}
L3 Collaboration, L3 Note 2568, presented at ICHEP 2000, Osaka, Japan.

\bibitem{opal}
OPAL Collaboration, OPAL Physics Notes PN456.


\bibitem{cdft}
M.~Cacciari, V.~Del Duca, S.~Frixione and Z.~Trocsanyi,
JHEP {\bf 0102}, 029 (2001)
[hep-ph/0011368].

\end{thebibliography}
\end{document}